# LUNG NODULE CLASSIFICATION ON CT SCAN PATCHES USING 3D CONVOLUTIONAL NEURAL NETWORKS

*Sydorskyi V. S.* – Postgraduate Student, Institute for Applied System Analysis
of National Technical University of Ukraine "Igor Sikorsky Kyiv Polytechnic Institute"
ORCID ID: 0000-0001-9697-7403

*Lung cancer remains one of the most common and deadliest forms of cancer worldwide. The likelihood of successful treatment depends strongly on the stage at which the disease is diagnosed. Therefore, early detection of lung cancer represents a critical medical challenge. However, this task poses significant difficulties for thoracic radiologists due to the large number of studies to review, the presence of multiple nodules within the lungs, and the small size of many nodules, which complicates visual assessment. Consequently, the development of automated systems that incorporate highly accurate and computationally efficient lung nodule detection and classification modules is essential.*

*This study introduces three methodological improvements for lung nodule classification: (1) an advanced CT scan cropping strategy that focuses the model on the target nodule while reducing computational cost; (2) target filtering techniques for removing noisy labels; (3) novel augmentation methods to improve model robustness. The integration of these techniques enables the development of a robust classification subsystem within a comprehensive Clinical Decision Support System for lung cancer detection, capable of operating across diverse acquisition protocols, scanner types, and upstream models (segmentation or detection).*

*Two task formulations are explored: multiclass and binary, along with several aggregation methods for converting multiclass predictions into binary outcomes. An extensive ablation study is conducted to quantify the contribution of each proposed methodological improvement.*

*The multiclass model achieved a Macro ROC AUC of 0.9176 and a Macro F1-score of 0.7658, while the binary model reached a Binary ROC AUC of 0.9383 and a Binary F1-score of 0.8668 on the LIDC-IDRI dataset. These results outperform several previously reported approaches and demonstrate state-of-the-art performance for this task.*

*Сидорський В. С. Класифікація легеневих утворів на фрагментах КТ-зображень із використанням тривимірних згорткових нейронних мереж*

*Рак легенів залишається одним із найпоширеніших і найсмертельніших видів раку у світі. Ймовірність успішного лікування значною мірою залежить від стадії, на якій було встановлено діагноз. Тому раннє виявлення раку легенів є надзвичайно важливим медичним завданням. Проте ця задача створює значні труднощі для торакальних радіологів через велику кількість досліджень, які необхідно проаналізувати, наявність множинних утворів у легенях і невеликі розміри багатьох із них, що ускладнює візуальну оцінку. У зв'язку з цим розробка автоматизованих систем, які містять високоточні та обчислювально ефективні модулі виявлення й класифікації легеневих утворів, є вкрай актуальною.*

*У цьому дослідженні представлено три методологічні вдосконалення для задачі класифікації легеневих утворів: (1) удосконалену стратегію фрагментування КТ-зображень, яка дозволяє моделі зосередитися на цільовому утворі та зменшити обчислювальні витрати; (2) методи фільтрації цільових міток для усунення зашумлених анотацій; (3) нові методи аугментації, спрямовані на підвищення стійкості моделі.*

*Інтеграція цих підходів дозволяє створити надійну підсистему класифікації у складі комплексної системи підтримки прийняття клінічних рішень для виявлення раку легенів, здатної працювати з різними протоколами проведення КТ дослідження, типами сканерів і вхідними моделями (сегментації чи детекції).*

*Розглянуто дві постановки задачі – багатокласову та бінарну, а також кілька методів агрегації для перетворення багатокласових прогнозів у бінарні. Проведено розширене дослідження для кількісної оцінки внеску кожного запропонованого методологічного вдосконалення.*

*Багатокласова модель досягла Macro ROC AUC = 0.9176 і Macro F1 = 0.7658, тоді як бінарна модель показала Binary ROC AUC = 0.9383 і Binary F1 = 0.8668 на наборі даних LIDC-IDRI. Отримані результати перевищують показники низки попередніх підходів і демонструють ефективність кращу від сучасного стану галузі для цієї задачі.*

***Ключові слова:*** *аналіз медичних зображень, згорткові нейронні мережі, класифікація раку легенів, система підтримки прийняття клінічних рішень, аугментація даних, глибинне навчання, LIDC-IDRI.*

**Introduction.** Lung cancer is one of the most common types of cancer worldwide. At the same time, it has one of the lowest survival rates, especially when diagnosed and treated at later stages [1]. These factors indicate that early-stage diagnosis of lung cancer is crucial for successful treatment. To address this challenge, many countries have launched national lung screening programs aimed at detecting high-risk groups (such as long-term smokers, patients with a history of chronic obstructive pulmonary disease, or those with a family history of lung cancer) and initiating treatment as early as possible [2, 3].

Most of these screening initiatives rely on low-dose computed tomography (LDCT) [3], followed by expert analysis performed by thoracic radiologists. The interpretation of each CT study is a highly labor-intensive process [4]: radiologists must first localize all suspicious nodules, which can easily be confused with other anatomical structures such as blood vessels, and then characterize each detected nodule according to the RECIST [5] framework. This workflow requires a substantial amount of human effort and is prone to observer variability and human error. When scaled to national screening programs involving thousands or even millions of patients, such manual analysis becomes practically infeasible without automated or semi-automated tools, such as Clinical Intelligent Decision Support Systems (CIDSS).

Decomposing the entire screening workflow, the main stages can be outlined as follows [6]:

1. Localization of lung nodules
2. Classification of each nodule by its level of malignancy suspicion
3. Characterization of each nodule
4. Mapping each case to the RECIST framework

This study focuses on exploring methods and models aimed at the second stage of this pipeline – classification of suspicious lung nodules. Recent approaches employ deep and machine learning techniques for this purpose. These include the use of 2D Convolutional Neural Networks (CNNs) on individual CT slices, 3D CNNs applied to full or cropped CT volumes, and classical machine learning models operating on extracted geometric and intensity-based features [7]. However, most existing approaches suffer from several limitations:

– Information loss, caused by relying on a limited number of axial slices, extreme downscaling of 3D data, or lossy feature extraction.

– High noise in target labels, introduced by human annotation errors and "borderline" cases, where the malignancy of a nodule cannot be clearly determined. Such ambiguity is typical for small or irregularly shaped nodules.

– Low robustness to localization inaccuracies produced by probabilistic detection models, as most methods assume perfect expert annotations and do not account for spatial shifts or uncertainty in the nodule position.

This work proposes a lung nodule classification method that:

– is based on 3D CNNs operating on high-dimensional CT patches, preserving full spatial information;

– employs extensive augmentation to improve generalization to localization shifts and imaging variations across scanners and acquisition settings;

– investigates strategies for noise reduction in the target variable to obtain a more stable and reliable classifier.

It is also important to note that the third step of the pipeline (Characterization of each nodule) may similarly benefit from machine or deep learning methods for assigning categorical descriptors to nodules (e.g., type, shape, or margin definition). Therefore, the models and techniques proposed in this work can be directly extended to address this stage as well.

**Related Work.** A typical approach to lung nodule classification is to locate the CT axial slice containing the largest cross-sectional diameter of the nodule and apply a classical 2D convolutional neural network (CNN) [8] such as ResNet [9], DenseNet [10], or EfficientNet [11]. This strategy can yield good results for large nodules that are clearly distinguishable from blood vessels or calcified regions.

A modification of this approach involves stacking several consecutive CT slices channel-wise and feeding them into a 2D CNN [8, 12]. This method allows the model to incorporate limited contextual information along the third (z-) axis, thereby improving precision. However, because the convolutional kernels remain two-dimensional, such models cannot fully exploit the available volumetric information. Moreover, these methods typically include only a few adjacent axial slices, which restricts the receptive field along the depth dimension.

Another extension is the use of hybrid architectures that combine CNNs with sequential models such as RNNs, GRUs, LSTMs, or Transformers [8]. In these approaches, CNNs first extract spatial feature maps from each slice, and the sequence model aggregates information across slices. This design enables learning of deeper dependencies between axial planes and typically achieves higher accuracy, particularly for nodules extending across multiple slices. Nonetheless, due to the two-level structure, the CNN component still lacks complete 3D spatial awareness, while the sequential component is limited in modeling fine-grained spatial relationships within each slice.

The most natural solution for lung nodule classification in 3D CT scans is to employ 3D CNNs, which extend conventional 2D architectures (e.g., ResNet, DenseNet, EfficientNet) into three dimensions [8,13]. Using 3D convolutional kernels, these models can capture full volumetric context and spatial structure. Their primary limitations, however, are high computational requirements and data scarcity. In most practical implementations, CIDSS are constrained by Graphical Processing Unit (GPU) memory (VRAM), making it infeasible to process entire 3D scans directly. Additionally, publicly available CT datasets are relatively small, typically containing only 2–3 thousand scans, and treating each scan as a single data sample results in limited training size. Combined with the much larger parameter count of 3D CNNs compared to 2D CNNs, this often leads to severe overfitting.

A practical way to apply 3D CNNs for lung nodule classification is the patch-based approach, where the full CT volume is divided into smaller 3D crops. This technique alleviates computational constraints, as patches are much smaller (e.g., a full scan of approximately [512, 512, 300] voxels versus a crop of [64, 64, 64]). If the system can generate accurate bounding boxes and the training dataset provides corresponding annotations, such patches retain all relevant information, since most nodules are significantly smaller than the patch size. Moreover, each scan may contain multiple nodules, effectively enlarging the training set. Despite these advantages, 3D CNNs trained on cropped patches still face challenges related to inaccurate bounding boxes, noisy labels, discrepancies between training and predicted bounding boxes, and variations in CT scanner configurations, all of which can affect the consistency and quality of input data.

**Method.**

**Dataset Preparation.** To standardize the input data (3D CT patches) and reduce noise in the target variable annotations, the following preprocessing steps were applied:

– Resampling of input volumes to isotropic voxel spacing and slice thickness.

– Normalization of intensity values.

– Aggregation of annotations across several radiologists, with exclusion of highly noisy categories.

The algorithm is designed to operate on low-dose CT scans both with and without contrast enhancement. However, the distribution of voxel geometry across the dataset varies substantially: pixel spacing (physical size in the X–Y plane) ranges from 0.461 mm to 0.977 mm, and slice thickness (physical size in the Z plane) varies from 0.45 mm to 5.0 mm. To standardize these parameters for convolutional operations, a representative near-median value was selected across the training data: 0.625 mm × 0.625 mm for pixel spacing and 1.0 mm for slice thickness.

Another critical characteristic is the intensity of CT voxels, which may differ due to contrast usage, scanner manufacturer (GE Medical Systems, Siemens, Philips, Toshiba), and individual acquisition protocols. Intensity normalization was performed using equation (1) with the following parameters: $a_{min} = -1024, a_{max} = 700, b_{min} = 0, b_{max} = 1$. This rescaled the original intensity range of [−1024,700] Hounsfield Units (HU)–covering most anatomical structures within the thoracic region–into the normalized range [0,1], which is standard for CNN inputs.

$$I_{out} = b_{min} + \frac{(I_{in} - a_{min}) \cdot (b_{max} - b_{min})}{a_{max} - a_{min}}, \tag{1}$$

The target variable represents the level of malignancy suspicion for each nodule, categorized into five classes: Highly Unlikely, Moderately Unlikely, Indeterminate, Moderately Suspicious, and Highly Suspicious. Each nodule was annotated by up to four radiologists (1–4 annotations per case, median = 3). A common approach is to use the highest suspicion level across annotators to reflect the worst-case scenario. However, this approach often introduces label noise; therefore, the classes were mapped to an ordinal numerical scale: $HighlyUnlikely \rightarrow 0$, … , $HighlySuspicious \rightarrow 4$, and the median label per nodule was subsequently taken. Since the Indeterminate category typically corresponds to uncertain or inconclusive cases where radiologists cannot make a reliable judgment, all nodules labeled as Indeterminate were excluded from the dataset to further reduce noise. Finally, the task can be reformulated as a binary classification problem using the following mapping:

– *HighlyUnlikely* → *NotDangerous*

– *ModeratelyUnlikely* → *NotDangerous*

– *ModeratelySuspicious* → *Dangerous*

– *HighlySuspicious* → *Dangerous*

The model will then be optimized as a binary classifier, with the predicted probability of the "Dangerous" class interpreted as the level of malignancy suspicion for each nodule.

**Crop Selection.** To generate standardized 3D input patches, the following crop selection strategy was applied:

1. The center of the crop is aligned with the center of the nodule bounding box.

2. The nodule bounding box is then fitted into a fixed-size cubic crop of [64,64,64] voxels. Two scenarios are possible:

a. If the bounding box is equal to or smaller than the fixed crop along all axes (which occurs in approximately 98.8% of cases), the bounding box is placed in the center of the crop.

b. If the bounding box exceeds the crop size along at least one axis, a padding of 8 voxels per side is applied along each "larger" axis. The resulting extended crop is then resized to the fixed [64, 64, 64] shape.

3. When a bounding box is located near the boundary of the CT volume, all missing voxels required to complete the fixed-size crop are zero-padded.

Additionally, to help the CNN focus on the nodule region (since the crop is usually larger than the bounding box itself), an auxiliary input channel was introduced in the form of a bounding box mask applied to the same spatial region. The combined input is expressed by equation (2):

$$I = concat(X, M) \epsilon R^{2 \times H \times W \times D} ,\qquad (2)$$

where $X \epsilon R^{H \times W \times D} - 3D\,scan\,crop$

$M(x,y,z) = \{X(x,y,z)\,if\,(x,y,z) \in bounding\,box\,region, 0\,otherwise\}$ – bounding box masked region

An example is illustrated in Figure 1.

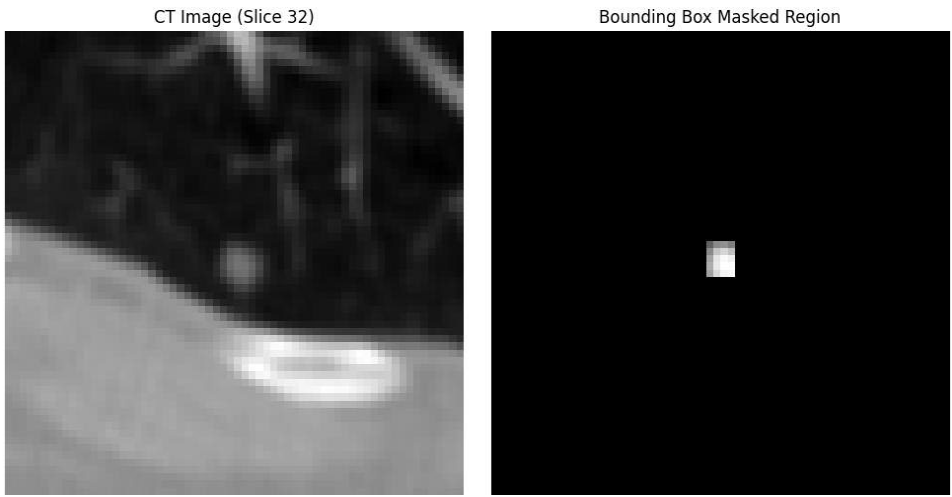

CT Image (Slice 32)          Bounding Box Masked Region

*Fig. 1. Axial Slice of 3D scan crop (left image) and bounding box masked region (right image)*

**Augmentations.** To improve the model's robustness to variations introduced by different scanners, pixel intensity deviations, acquisition protocols, and potential localization errors from upstream detection models, a series of data augmentations was applied. All augmentations were performed "on the fly" during training.

To adapt the model to possible shifts in the predicted bounding box, a Box Jittering augmentation was employed. The proposed cropping strategy assumes that the nodule is located at the center of the crop; however, in real-world scenarios, detection models may produce localization errors. Let the original bounding box be represented as: $B = [x_{min}, y_{min}, z_{min}, x_{max}, y_{max}, z_{max}]$, and the corresponding lengths along each spatial dimension as: $L = \left[ L_x, L_y, L_z \right] = \left[ x_{max} - x_{min}, y_{max} - y_{min}, z_{max} - z_{min} \right]$, that proposed augmentation is described by equation (3):

$$B' = \left[ x_{min} + \Delta_x, y_{min} + \Delta_y, z_{min} + \Delta_z, x_{max} + \Delta_x, y_{max} + \Delta_y, z_{max} + \Delta_z \right],$$

$$\text{where } \Delta_i \sim U \left( L_i \cdot \alpha_i^{min}, L_i \cdot \alpha_i^{max} \right) \tag{3}$$

and the resulting box $B'$ is constrained to remain within the original image bounds: $B' \subseteq [0, H] \times [0, W] \times [0, D]$. Given that the median bounding box side length across the dataset is relatively small (10.96 voxels), the jittering parameters were set to $\alpha^{min} = -0.75$ and $\alpha^{max} = 0.75$ for all spatial axes. An example of this augmentation is shown in Figure 2.

For the proposed two-channel cropping strategy, an additional augmentation was introduced: second-channel dropout with a probability of 0.5. This transformation randomly zeroes out the bounding box mask channel $M(x, y, z)$ to prevent the CNN from overfitting to it and encourage greater reliance on the image channel $X(x, y, z)$.

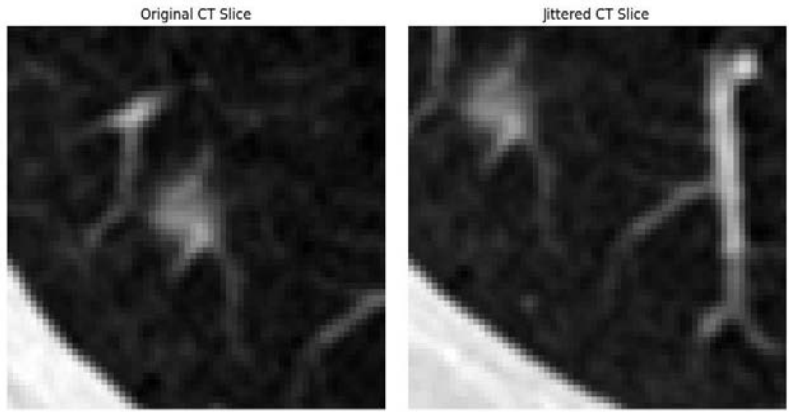

*Fig. 2. Original axial slice of 3D scan crop (left image) and same crop but jittered in X and Y axes (right image)*

In addition to the newly introduced augmentations, a set of standard 3D transformations was incorporated into the pipeline: random flips across all spatial axes, random 90° rotations in the X–Y plane, random zoom, random Gaussian noise, and random Gaussian smoothing.

**CNN Model, Optimization, Inference.** For all experiments, a ResNet-50 architecture was selected, following the recommendations outlined in [13].

The network's first convolutional layer was modified to have a stride of 1 and a kernel size of 3, with the number of input channels set to 2 to accommodate both the CT crop and an additional input channel. The residual shortcut type "B" was used for skip connections.

Model optimization was performed using the Adam optimizer [14] with a cosine learning rate schedule [15] under a one-cycle policy, where the learning rate gradually decreased from $1 \times 10^{-3}$ to $1 \times 10^{-5}$. For the multiclass formulation, the Cross-Entropy Loss function was applied, while for the binary classification version, the Binary Cross-Entropy Loss was used. Training was conducted for a maximum of 75 epochs with a batch size of 64. Each epoch consisted of approximately 16 steps. To mitigate overfitting, early stopping was applied based on the macro F1-score (for the multiclass setup) or binary F1-score (for the binary setup) computed on the validation set.

For inference on validation and test data, the best-performing model, as determined by the highest validation F1-score, was selected. To further improve prediction stability, Test-Time Augmentation (TTA) was employed following the approach of [16]. Predictions were generated for all possible flip combinations across the three spatial axes (a total of eight variants), and the final prediction for each sample was obtained by averaging the model outputs across these augmented views.

**Experiments.**

**Dataset and Validation.** For all experiments, the LIDC-IDRI dataset [17] was used. It contains 1,018 CT scans from 1,010 unique patients. After filtering out corrupted scans and cases without valid annotations, 850 scans were retained, comprising a total of 2,525 nodules distributed as follows:

– Indeterminate - 1177
– Moderately Unlikely - 532
– Moderately Suspicious - 332
– Highly Unlikely - 312
– Highly Suspicious - 172

In the case of using the aggregation method proposed in the Dataset Preparation Subsection.

To ensure reliable evaluation, given the limited dataset size and the presence of multiple nodules per scan (and per patient), a Grouped Cross-Validation strategy was employed. The data were split into five folds, ensuring that nodules from the same patient never appeared in both training and validation sets. For each experiment, five CNN models were trained, each using four folds for training and one fold for validation. Predictions from the five validation folds were then aggregated to obtain results for the entire dataset.

**Quantitative Results.** Three multiclass experiments were conducted:

– Multiclass Model 1: trained without the Indeterminate class and with Box Jittering augmentation.
– Multiclass Model 2: trained with the Indeterminate class and with Box Jittering augmentation.
– Multiclass Model 3: trained without the Indeterminate class and without Box Jittering augmentation.

All models were evaluated using macro-averaged metrics: ROC AUC, Recall, Precision, and F1-score. The results are presented in Table 1.

Table 1

**Experiment results of Multiclass models**

| Model Name | Macro Roc Auc ↑ | Macro Recall ↑ | Macro Precision ↑ | Macro F1 ↑ |
|---|---|---|---|---|
| Multiclass Model 1 | **0.9011** | **0.7333** | **0.7409** | **0.7346** |
| Multiclass Model 2 | 0.8255 | 0.5560 | 0.5560 | 0.5392 |
| Multiclass Model 3 | 0.8961 | 0.7213 | 0.7346 | 0.7249 |

Multiclass Model 1, which incorporates all proposed preprocessing and augmentation strategies, outperformed all others. The largest improvement was achieved by excluding the Indeterminate class (+7.5 ROC AUC), while Box Jittering provided a smaller but consistent gain (+0.5 ROC AUC).

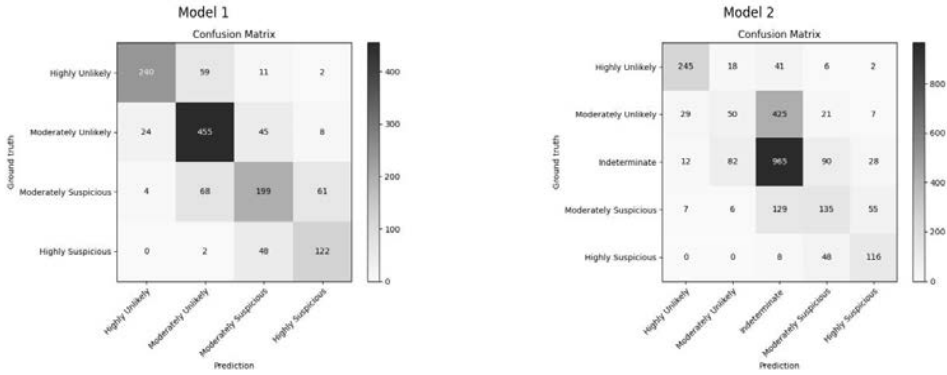

*Fig. 3. Confusion matrices of Multiclass Model 1 (right image) and Multiclass Model 2 (left image)*

Analysis of the confusion matrices in Figure 3 indicates that the Indeterminate class introduces substantial confusion–particularly between the Moderately Unlikely and Moderately Suspicious categories. Additionally, models trained with the Indeterminate class performed worse in identifying Highly Suspicious nodules, despite slightly better performance on Highly Unlikely cases. These findings confirm the hypothesis outlined in the Methods section regarding the detrimental effect of noisy Indeterminate labels.

Next, classification performance was assessed using bounding boxes generated by the nnU-Net-based segmentation [18] and localization module. Results are shown in Table 2.

Table 2

**Experiment results of Multiclass models on upstream model bounding boxes**

| Model Name | Macro Roc Auc ↑ | Macro Recall ↑ | Macro Precision ↑ | Macro F1 ↑ |
|---|---|---|---|---|
| Multiclass Model 1 | **0.9176** | **0.7644** | 0.7681 | **0.7658** |
| Multiclass Model 3 | 0.9173 | 0.7635 | **0.7685** | 0.7655 |

Performance on predicted bounding boxes is nearly identical across models with and without jittering. This is likely due to the upstream model's difficulty in detecting the most challenging nodules–these are thus absent from evaluation. Nevertheless, given that jittering enhances robustness under real-world conditions, Multiclass Model 1 remains the preferred configuration.

For binary classification, two models were trained, and multiclass models were also converted using probability aggregation: maximum (equation (4)) and sum (equation (5)). For both of these options, the probability of the negative class is defined by equation (6).

$$Prob_{Dangerous} = max\left(Prob_{ModeratelySuspicious}, Prob_{HighlySuspicious}\right),$$ (4)

$$Prob_{Dangerous} = Prob_{ModeratelySuspicious} + Prob_{HighlySuspicious},$$ (5)

$$Prob_{NotDangerous} = 1 - Prob_{Dangerous}.$$ (6)

Binary models:
– Binary Model 1: without Indeterminate class, with jittering.

– Binary Model 3: without Indeterminate class, without jittering.

Multiclass-derived models were denoted with Sum or Max prefixes, based on the aggregation method. Results are summarized in Table 3.

**Experiment results of Binary and aggregated Multiclass models**

| Model Name | Binary Roc Auc ↑ | Binary Recall ↑ | Binary Precision ↑ | Binary F1 ↑ |
|---|---|---|---|---|
| Binary Model 1 | **0.9383** | **0.8651** | 0.8685 | **0.8668** |
| Binary Model 3 | 0.9297 | 0.8373 | 0.8612 | 0.8491 |
| Multiclass Model 1 Sum | 0.9365 | 0.8552 | 0.8637 | 0.8594 |
| Multiclass Model 1 Max | 0.93 | 0.7599 | 0.8764 | 0.814 |
| Multiclass Model 2 Sum | 0.9273 | 0.8075 | 0.8772 | 0.8409 |
| Multiclass Model 2 Max | 0.9221 | 0.7341 | **0.8937** | 0.8061 |

Binary Model 1 achieved the highest overall performance, consistent with the multiclass results. Jittering again contributed to better performance, particularly in F1 and AUC metrics. Among the aggregated multiclass variants, the Sum aggregation approach outperformed the Max aggregation. This is intuitive, as multiclass outputs are normalized via a softmax function, meaning the sum of probabilities for the Dangerous classes preserves total probability mass. In contrast, the maximum operation reduces one component, slightly increasing Not Dangerous probability–improving precision but lowering recall.

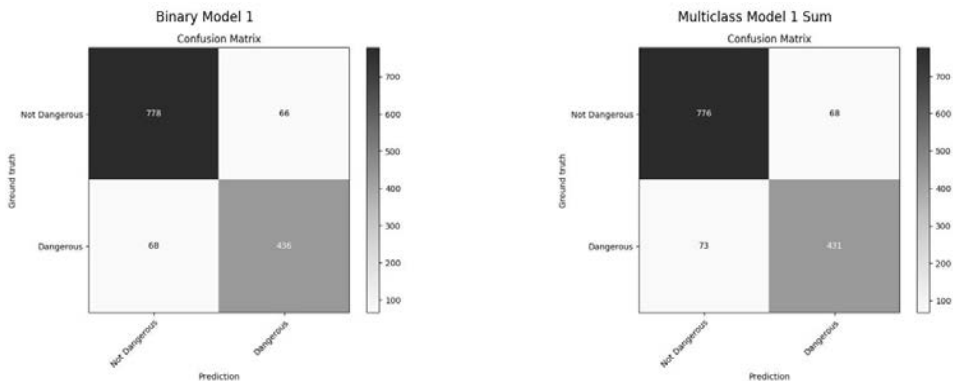

*Fig. 4. Confusion matrices of Binary Model 1 (right image) and Multiclass Model 1 Sum (left image)*

Analyzing the confusion matrices from Figure 4, there is no big difference in the performance of Binary Model 1 and Multiclass Model 1 Sum. Additionally, this small difference is evenly distributed between first-order and second-order errors.

**Visual Results.** Figure 5 illustrates qualitative examples of predictions from Binary Model 1 and Multiclass Model 1.

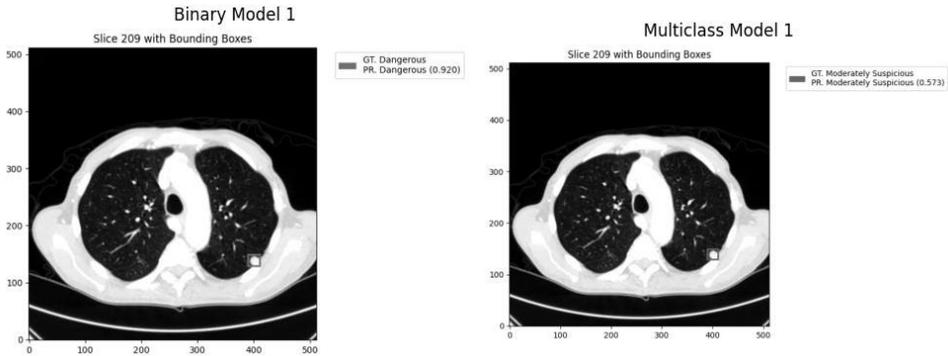

*Fig. 5. Visual results of Binary Model 1 (right image) and Multiclass Model 1 (left image)*

These results demonstrate how the proposed system can visualize malignancy predictions for each detected nodule in real clinical conditions.

**Comparison with Previous Work.** The proposed method was compared against several recent approaches on the LIDC-IDRI dataset. Results are summarized in Table 4.

Table 4

### Comparison with previous works

| Method | Roc Auc ↑ |
|---|---|
| Attention pyramid pooling network for artificial diagnosis on pulmonary nodules, Wang et al. (2024) [19] | 0.914 |
| Lung Nodule Classification Using Biomarkers, Volumetric Radiomics, and 3D CNNs, Mehta et al. (2019) [20] | 0.8 |
| Proposed Binary Model 1 | 0.9383 |
| Proposed Multiclass Model 1 | 0.9176 |

The proposed method, both in its binary and multiclass formulations, outperforms previously published CNN-based and attention-based architectures. These results indicate that the developed models provide a competitive and robust solution for the task of lung nodule malignancy classification.

**Conclusions and Limitations.** This study introduced a set of methodological improvements for lung nodule malignancy classification aimed at enhancing model accuracy, robustness, and computational efficiency. The proposed approach was designed to operate effectively across diverse clinical conditions and to be integrated as a component of a Clinical Intelligent Decision Support System (CIDSS). The main methodological contributions include:

– A cropping strategy with a masked auxiliary input, which emphasizes the most relevant region of interest (ROI). This enables the model to focus on diagnostically significant areas, improving accuracy while reducing computational cost.

– Bounding box jittering augmentation, which enhances robustness to localization errors propagated from upstream detection models and simultaneously contributes to improved classification performance.

– Pruning of noisy target categories, particularly the Indeterminate class, which reduces label ambiguity and improves model stability.

– Introduction and comparison of binary and multiclass model formulations, providing a systematic analysis of task framing and aggregation strategies.

By combining these elements, the final models achieved 0.9176 Macro ROC AUC and 0.7658 Macro F1 in the multiclass setting, and 0.9383 Binary ROC AUC and 0.8668 Macro F1 in the binary setting. These results demonstrate that the proposed methods substantially improve the reliability and interpretability of CNN-based lung nodule classification.

Despite the promising results, several limitations remain:

– Dataset diversity: All experiments were conducted using the LIDC-IDRI dataset, which may limit the generalizability of findings to data from other clinical centers or acquisition protocols.

– Model diversity: Only a single CNN backbone (ResNet-50) was evaluated. To confirm the general applicability of the proposed methods, further experiments with different architectures (e.g., DenseNet, EfficientNet, 3D Vision Transformers) are required.

– Annotation noise: Additional strategies for reducing annotation uncertainty should be explored, such as filtering very small nodules or those with minimal axial diameter, which often lead to labeling inconsistencies.

– Contextual reasoning: The current formulation classifies each nodule independently, disregarding potential correlations among multiple nodules within a single patient. Attention-based or relational models could be employed to capture such dependencies.

– Auxiliary signals: Incorporation of additional metadata or auxiliary learning objectives - either human-annotated or algorithmically derived [7] - may help guide the model toward more clinically consistent decision boundaries.

Future research will focus on:

– Evaluating the proposed methods on additional datasets and architectures to validate robustness and generalizability.

– Extending the current pipeline with advanced domain-specific augmentations and refined label filtering rules.

– Integrating attention-based mechanisms for context-aware pooling of multiple nodules and exploring the use of patient-level metadata to enhance clinical interpretability.

In summary, this work presents an improved and computationally efficient approach for lung cancer classification based on localized nodules. The developed methods represent a significant step toward building a comprehensive clinical decision support framework, capable of reducing radiologists' workload while improving diagnostic accuracy for one of the most prevalent and lethal forms of cancer.

**Acknowledgment.** The author expresses sincere gratitude to the Better Medicine OÜ team for their valuable medical and technical insights, which were instrumental in the preparation of this scientific paper. Special thanks are extended to the Armed Forces of Ukraine, whose resilience and dedication made it possible to carry out this work in safety.

During the preparation of this manuscript, the author utilized OpenAI's ChatGPT to assist with improving clarity, coherence, and formatting. All generated content was thoroughly reviewed and edited, and the author assumes full responsibility for the final version of this publication.